\documentclass[journal]{IEEEtran}
\usepackage{cite}
\usepackage{amsmath,amssymb,amsfonts}
\usepackage{graphicx}
\usepackage{textcomp}
\usepackage{bm}
\usepackage{xcolor}

\newif\ifcomments
\commentsfalse   

\ifcomments

\newcommand\MAD[1]{{\color{purple}[MAD: \textit{#1}]}}
\else

\newcommand\MAD[1]{}
\fi

\def\BibTeX{{\rm B\kern-.05em{\sc i\kern-.025em b}\kern-.08em
		T\kern-.1667em\lower.7ex\hbox{E}\kern-.125emX}}

\begin{document}

\title{Arithmetic Reconciliation for CVQKD: Challenges and Feasibility}

\author{Rávilla R. S. Leite, Juliana M. de Assis, Micael A. Dias and Francisco M. de Assis
	\thanks{Rávilla R. S. Leite and Francisco M. de Assis, Department of Electrical Engineering, Federal University of Campina Grande (UFCG), Campina Grande-PB, Brazil. E-mails: ravilla.leite@ee.ufcg.edu.br, fmarcos@dee.ufcg.edu.br. This work was partially financed by CNPq. (311680/2022-4 and 140827/2022-6)}
	\thanks{Juliana M. de Assis, Department of Statistics, Center for Exact and Natural Sciences (CCEN), Federal University of Pernambuco (UFPE), Recife-PE, , Brazil (e-mail: juliana.massis@ufpe.br)}
	\thanks{and Micael A. Dias, Department of Electrical and Photonics Engineering, Technical University of Denmark, Lyngby, Denmark, (mandi@dtu.dk)}
	\thanks{This work was supported in part by CNPq (311680/2022-4 and 140827/2022-6), by Fundação de Amparo à Ciência e Tecnologia do Estado de Pernambuco - FACEPE (APQ-1341-1.02/22), by CAPES (88887.014949/2024-00), and by the European Union (HORIZON-MSCA-2023 Postdoctoral Fellowship, 101153602 - COCoVaQ).}}

\maketitle

\begin{abstract}
Continuous variable quantum key distribution allows two legitimate parties to share a common secret key and encompasses reconciliation protocols. A relatively new reconciliation protocol, Arithmetic Reconciliation, presents low complexity and has increasing reconciliation efficiency with lower SNRs. In this paper, we obtain reconciliation efficiencies for this protocol in realistic scenarios, by means of estimation of mutual information, and we also present rates for sequence match of secret keys by Alice and Bob. Results show that this technique is feasible and promising to continuous variable quantum key distribution applications. 
\end{abstract}



\maketitle

\section{Introduction}
\label{sec:introduction}

Advances in quantum computing compose a threat to secure data transmission, since quantum computers may factorize large prime numbers \cite{shor1994algorithms}, weakening the security of encryption methods, among other capabilities. One way to circumvent this issue may be the use of quantum secret keys in the one-time pad system \cite{shannon1949communication, BB84}. The physical transport of large sequences of secret bits can be inneficient and laborious, but this may be overcome by quantum key distribution (QKD).

QKD protocols enable two legitimate parties, Alice and Bob, to achieve a common secret  binary sequence through a quantum system. The secrecy of this common binary sequence is due to the no-cloning theorem of quantum mechanics, since any measurement from an eavesdropper (Eve) would disturb the quantum state \cite{wootters1982single}. QKD protocols may be performed with discrete variables \cite{BB84, E91, B92}, such as the polarization or phase of single photons or with continuous variables, such as the quadratures of quantized electromagnetic fields \cite{GG02, GG2003, Wedbrook2004switching, li2017continuous, Leverrier2019asymptotic}. 

Continuous-variable quantum key distribution (CVQKD) presents some advantages over its discrete variable counterpart (DVQKD). It is relatively easy to implement with the existing telecommunications equipments and allows higher secret rates \cite{van2004reconciliation, jouguet2011long, jouguet2014high, Leverrier2019asymptotic}. Moreover, DVQKD requires cryogenic temperatures for single photon detection, while CVQKD may be used with room temperatures \cite{milicevic2017key}. These advantages are the main reasons why scientific literature refers more CVQKD than DVQKD.

CVQKD is composed by a quantum information transmission and a post-processing of data through a classical public and authenticated channel \cite{zhu2025novel}. Briefly speaking, in quantum information transmission, Alice and Bob acquire correlated Gaussian measurements. The classical post-processing stage encompasses information reconciliation, where the correlated continuous values of Alice and Bob will be transformed in a common (discrete) secret key, and privacy amplification, which destroys any information Eve may have about the secret key. In direct reconciliation (DR), Bob will correct his sequence to match Alice's, whereas in reverse reconciliation (RR), it is Alice who will correct her sequence to match Bob's. Surprisingly, the direction of reconciliation is important in the papers involving the subject until now, and RR presents potentially a higher efficiency for secret key generation \cite{origlia2025soft}, since it allows the protocol to be executed even on channels with losses greater than 3 dB \cite{GG_virtual_entanglement}. The most wide spread reconcilition methods are Slice Error Correction (SEC) and Multidimensional Reconciliation (MD) \cite{van2004reconciliation, liu2024road, Leverrier_2008}. SEC has better performance in applications of short distances, whereas MD reconciliation has improved performance at signal-to-noise ratio (SNR) lower than 0 dB \cite{jia2024high, Leverrier_2008}.

A less commonly discussed information reconciliation scheme is Arithmetic Reconciliation (AR) \cite{Laryssa:SbRT2018, MicaelFMA:2024}. Originally proposed by Araújo, Assis, and Albert \cite{Laryssa:SbRT2018}, this technique is inspired by elements of Information Theory, such as Shannon–Fano–Elias coding and Arithmetic Coding. AR maps each realization of a continuous random variable to its Cumulative Distribution Function (CDF), thereby projecting it onto the unit interval $[0,1]$. This mapping simplifies the quantization process and enables a binary representation of the variable through its binary expansion. Because Arithmetic Coding ensures that the binary expansion of a continuous value uniformly distributed over the unit interval yields i.i.d. Bernoulli(1/2) bits, the resulting bit strings are statistically well-behaved for key extraction.

The main advantage of AR is its reduced computational complexity compared to methods such as Slice Error Correction (SEC) and Multidimensional Reconciliation (MD). It requires neither sophisticated decoding procedures nor external random number generators, since it leverages the intrinsic randomness of quantum measurements for key generation. Simulations reported in \cite{MicaelFMA:2024} show that AR achieves reconciliation efficiency above $90\%$ at an SNR below of $-3.6$ dB, with comparable performance in both direct reconciliation (DR) and reverse reconciliation (RR) settings. More recently, \cite{ravilla2025arithmetic} extended some results on AR, providing additional analytical insights and by means of simulations.

Despite the advances in the above cited references concerning the AR protocol, the scientific literature lacks a complete explanation of the procedure, specially in the last step where the secret key acquisition for both Alice and Bob happens, by means of a binary error correction protocol, such as Low Density Parity Check (LDPC) codes. The main objective of this paper is to fill this gap by providing analytical and computational results to back up the feasability of AR. By parametrizing Alice and Bob's correlated data with the effective signal-to-noise ratio, we show that mutual information is preserved under the Distributional Transform Expansion (DTE) and derive expressions for the binary symetric channel (BSC) induced by the expansion method. The quantization efficiency is estimated for the case of binary-input-continuous-output induced channels where soft-information is kept for the decoder, allowing estimated efficiency above 0.95 at low SNR. We also show simulation results with the entire reconciliation procedure by using an LDPC code at SNR from 2 to 7 dB (according to the code rate provided).

Beyond the introductory section, the rest of the paper is organized as follows: Section \ref{sec:methods} explains AR procedures, Section \ref{sec:results} brings simulation results and Section \ref{sec:conclusion} concludes the paper. Some calculations concerning channel performance are left to Appendix \ref{appendix:ber}. 

\subsection{Notation}
In order to achieve this objective, firstly, we present some notation. We use capital letters to denote random variables, and lowercase letters to denote the specific values assumed by them. Subscripts in random variables denote a subchannel, that is, a binary random variable that is obtained by the DTE performed after Alice's and Bob's measurements, as we will clarify in the following section. The operators $E()$ and Var$()$ denote expected value and variance, respectively. The following well known functions will be frequently called:
\begin{equation}
	Q(x) \equiv \int_x^\infty \frac{1}{\sqrt{2\pi}} e^{\frac{-x^2}{2}} dx,   \ \ \ \ \ \ 
	\Phi(x)\equiv \int_{-\infty}^x \frac{1}{\sqrt{2\pi}} e^{\frac{-x^2}{2}} dx. \nonumber
\end{equation}

\section{Arithmetic Reconciliation}
\label{sec:methods}

AR consists of two stages. First, the continuous variables obtained by Alice and Bob after the quantum stage are quantized. When considering protocols with Gaussian modulation of coherent states, the GG02 and the "no-switching" protocols share the same structure, differing mainly on the measurement applied, where the former uses homodyne detection and the later, heterodyne \cite{GG02, GG2003, Wedbrook2004switching}. When homodyne detection is used, Bob must announce the active choice of measurement basis so that Alice can discard the mismatching quadratures. This stage is named sifting. Besides the measurement used, both strategies will result on Alice and Bob sharing correlated Gaussian random variables, up to a correction factor on the variance of Bob's outcomes \cite{MicaelFMA:2024, Laudenbach_2018}.

The quantization technique\footnote{Here, the term “quantization” refers to its use in communication theory, namely the process of mapping a continuous-amplitude variable onto a discrete set of values — in our case, binary outcomes.} employed is based on an well-known and very useful statement, refereed here as Lemma of Arithmetic Coding, from which the protocol derives its name, which states that “\textit{transforming a random variable with a continuous distribution function by its cumulative distribution function will always lead to a random variable uniformly distributed in the interval} $[0,1]$” \cite[p.437]{cover2006}. After this mapping to the unit interval, the binary expansion of the obtained values is performed, transforming each initially measured value into binary sequences with $m$ bits of precision. The bits obtained with the binary expansion of variables contained in the unit interval are independent and Bernoulli $(1/2)$. This Lemma is known in Copula Theory as Distributional Transform, and for this reason, the quantization technique was named Distributional Transform Expansion by Dias and Assis \cite{MicaelFMA:2024}.


In the second stage, a reverse reconciliation scheme is always considered. Here, Alice and Bob's sequences do not undergo traditional channel coding, in which an error-correcting code is used to generate codewords that must be recovered at the receiver, but rather syndrome coding (or Slepian-Wolf coding) is applied. Thus, Alice and Bob's binary sequences are random i.i.d. sequences belonging to $GF(2)^n$, which will be treated as correlated sources, to which LDPC codes will be applied to perform the syndrome calculation (size $n-k$). The syndrome calculation compresses Bob's source (RR), transforming its $n$-bit sequence into $n-k$ bits.

\begin{figure*}
\begin{center}
\includegraphics[width=0.8\textwidth]{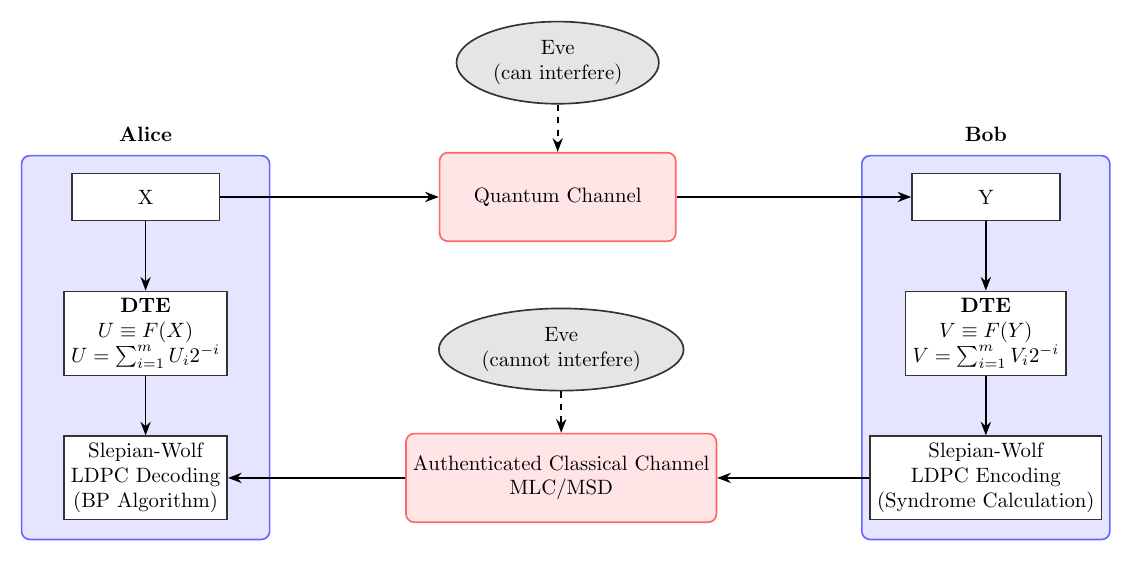}
\caption{Sketch of the complete AR scheme.}
\label{fig:AR}
\end{center}
\end{figure*}

The Slepian-Wolf Theorem shows that by compressing Bob's source at a rate up to $H(Y|X)$, Alice can reconstruct it using her sequence $X$ as side information in a joint decoder \cite{slepian1973coding}. Liveris et al. \cite{Liveris_2002} address the application of LDPC codes in distributed source coding situations and present a modification to the sum-product algorithm to include the syndrome in decoding.Fig. \ref{fig:AR} illustrates the complete AR scheme.


Following \cite{jouguet2014high, jouguet2011long, Leverrier2019asymptotic}, the quantum channel was interpreted as an Additive White Gaussian Noise (AWGN) channel, so, after quantum information transmission, Alice and Bob have correlated Gaussian values, $X$ and $Y$, respectively. Considering a normalization, we may write, without loss of generality, $Y = X+\frac{1}{\sqrt{SNR}}Z$, and $X, Z \sim \mathcal{N}(0,1)$, where $X$ and $Z$ are independent. Since $X$ and $Y$ are independent Gaussian variables, and $Y$ is a linear combination of them, $Y$ will also be a Gaussian variable, with zero mean and variance $\mbox{Var}(Y) = 1+\frac{1}{SNR}$.

The correlation between $X$ and $Y$ may be expressed in terms of their $SNR$ as:
\begin{align}
\rho_{XY} &= \dfrac{E(XY) - E(X)E(Y)}{\sqrt{\mbox{Var}(X)\mbox{Var}(Y)}} \nonumber \\
&= \dfrac{E(X(X+\frac{1}{\sqrt{SNR}}Z))}{\sqrt{1+\frac{1}{SNR}}} \nonumber \\
&= \dfrac{1}{\sqrt{1+\frac{1}{SNR}}} \label{eq:corr}
\end{align}

The quantum information transmission will be performed $n$ times in AR scheme, so Alice will have an i.i.d sequence of realizations of $X$ and Bob will have an i.i.d sequence of realizations of $Y$, and each pair $(X, Y)$, will be correlated according to (\ref{eq:corr}). 

\subsection{Distributional Transform Expansion}

\begin{figure}
\centering
\includegraphics[width=0.45\textwidth]{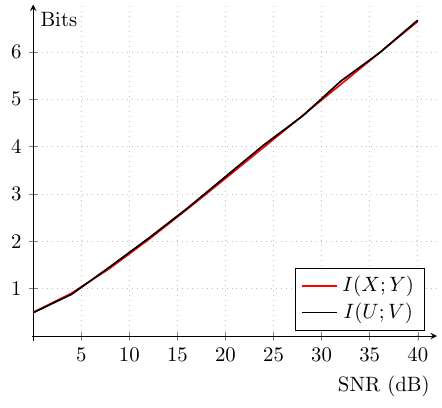}
\caption{Analytical mutual information between $X$ and $Y$, and estimates of $I(U;V)$, for different SNRs.}
\label{fig:infoXYandUV}
\end{figure}

As mentioned above, the DTE application to the Gaussian realizations of Alice and Bob is the first step in AR scheme. The DTE will firstly compute the cumulative distribution function of each Gaussian realization of Alice, $U = F_X(X) = \Phi(X)$, and Bob, $V = F_Y(Y) = \Phi(Y/\sqrt{1+\frac{1}{SNR}})$. Interestingly, this transformation preserves information, because mutual information is invariant under homeomorphisms \cite{sisak2024mutual}.  Fig. \ref{fig:infoXYandUV} presents estimates of $I(U;V)$ (using KSG estimator \cite{kraskov2004estimating}), with sample size $n=5\cdot 10^3$, as also the analytical value of $I(X;Y) = \dfrac{1}{2}\log(1+SNR)$, considering different SNRs. 

Secondly, DTE will perform a binary expansion with $m$ bits precision of the values of $U$ and $V$. Specifically, a binary expansion with $m$ bits precision of a number $U \in [0,1]$ will generate a sequence of bits $0.U_{1}U_{2}\dots U_{m}$, such that \cite{micael:SbRT2024}:
\begin{align}
	U_{k} = \mathbb{I}\{ U \in \cup _{j=1}^{2^{k-1}}[2^{-k}\cdot (2j-1), 2^{-k}\cdot 2j) \}, \label{eq:bitseq}
\end{align}
where $\mathbb{I}\{A\}$ is the indicator function of an event $A$, which equals 1 when $A$ occurs and 0 otherwise. For instance, the binary expansion of a realization $U$ results in $U_{1}$ and $U_{2}$:
\begin{align}
	U_{1} &= \begin{cases}
		0,\, \mbox{ if } U<1/2,\\
		1,\, \mbox{ if } U\geq 1/2,
	\end{cases} \nonumber \\ 
	U_{2} &= \begin{cases}
		0,\, \mbox{ if } U\in[0,1/4)\cup[1/2,3/4),\\
		1,\, \mbox{ if } U\in[1/4,1/2)\cup[3/4, 1].
	\end{cases} \nonumber 
\end{align}

\begin{figure}
\centering
\includegraphics[width=0.45\textwidth]{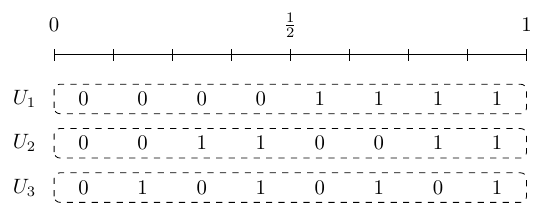}
\caption{Unit interval and bits assignment, $m=3$.}
\label{fig:unitinterval}
\end{figure}

Fig. \ref{fig:unitinterval} illustrates the partition of the unit interval and the corresponding bits with precision $m=3$.

\begin{figure}
\centering
\includegraphics[width=0.45\textwidth]{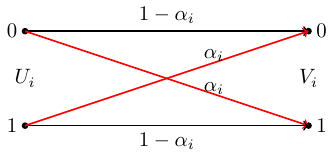}
\caption{BSC illustration in an AR scheme.}
\label{fig:bsc}
\end{figure}

At the end of each measurement and DTE procedure, Alice and Bob will each have a sequences of $m$ bits, $\{U_i\}$ and $\{V_i\}$, $i=1, \dots , m$, respectively. It is important to stress that the bits in the binary expansion are independent Bernoulli variables, with parameter $1/2$, since $U$ and $V$ are uniformly distributed \cite{cover2006}. However, each pair of bits $(U_{i}, V_{i})$ are correlated according to the development in (\ref{eq:rhoUV}):
\begin{align}
E(U_{i}) &= E(V_{i}) = 1/2 \nonumber \\
\mbox{Var}(U_{i}) &= \mbox{Var}(V_{i}) = 1/4 \nonumber \\
E(U_{i}V_{i}) &= 1.1.\mbox{Pr}(U_{i,(j)}=1,V_{i,(j)}=1)\nonumber \\ 
&= 1.1.\mbox{Pr}(V_{i}=1|U_{i}=1).\mbox{Pr}(U_{i}=1) = \frac{1}{2}(1-\alpha_{i})\nonumber \\
\rho _{U_{i}V_{i}} &= \dfrac{\frac{1}{2}(1-\alpha_{i}) - \frac{1}{4}}{1/4}\nonumber \\
&= 1-2\alpha_{i}, \label{eq:rhoUV}
\end{align}
where $\alpha_{i} = \mbox{Pr}(V_{i} = v|U_{i} = u)$, $u\neq v$, for $u,v\in \{0,1\}$, $i=1,2,\dots , m$. Thus, the relation between $U_{i}$ and $V_{i}$ may be interpreted as a binary symmetric channel (BSC, see Fig. \ref{fig:bsc}). Due to symmetry and the fact that the bits are Bernoulli(1/2), notice that $\alpha_{i} = \mbox{Pr}(V_{i}\neq U_{i})$. The calculation of probabilities $\alpha_{i}$ for different SNRs is partly presented in reference \cite{ravilla2025arithmetic}, Appendix \ref{appendix:ber} brings a thourough analysis, with $i=1,2$, for the sake of completeness.

\subsection{Reconciliation Efficiency for AR Scheme}

Reconciliation efficiency indicates how close to the Shannon limit the reconciliation algorithm can operate \cite{jouguet2014high, bloch_et_al_2006, milicevic2017key} and depends on the quantization efficiency and error correction code used in each channel. Quantization efficiency indicates how much of the mutual information $I(X;Y)$ has been preserved. In the AR context, in DR, the goal is to maximize $\sum_{i=1}^mI( U_i;Y)$, whereas in RR, $\sum_{i=1}^mI(V_i;X)$, with $\{U_i\}$ and $\{V_i\}$ being binary sequences and $X$ and $Y$ continuous. As shown in \cite{MicaelFMA:2024}, the reconciliation efficiency of DR and RR can be obtained respectively by:
\begin{equation}
	\label{eq:beta_dte}
	\beta_{q}^{\rightarrow} = \frac{\sum_{i=1}^{m}{I(U_i;Y)}}{I(X;Y)},
\end{equation}
\begin{equation}
	\beta_{q}^{\leftarrow} = \frac{\sum_{i=1}^{m}{I(V_i;X)}}{I(X;Y)}.
\end{equation}

AR treats CVQKD reconciliation protocol as a Slepian-Wolf problem of distributed source coding \cite{slepian1973coding}. Considering a RR scheme in which Alice has the Gaussian measurement $X$, which will be used as side information in the joint decoder, according to the Slepian-Wolf Theorem, she needs $H(V_i|X)$ bits to reconstruct Bob's component of the sequence, $V_i$. To maximize $I(V_i;X)$, Bob must compress his binary sequence at a rate $H(V_i|X)$ and send it through an authenticated classical channel, assumed error-free. In this work, source compression is performed by calculating the syndrome using LDPC codes \cite{bloch_et_al_2006, Liveris_2002}, called Syndrome Coding.

Reference \cite{Ravilla_sbrt_2024} shows that when the LDPC code used to calculate the syndrome has a rate compatible with the capacity of the considered channel, the decoding algorithm can recover the desired sequence, transforming our problem into a channel coding problem. Thus, we can consider a multilevel coding (MLC) scheme, where each subchannel obtained with the DTE can be coded independently using an LDPC code with a rate compatible with its capacity, i.e. $R_i \leq C_i$ for $1 \leq i \leq m$ \cite{van2004reconciliation, bloch_et_al_2006, MicaelFMA:2024}. The maximum efficiency is achieved when $R_i = C_i$.

The Arithmetic Coding Lemma guarantees that $F_Y(Y)$ has a uniform distribution in the interval $[0,1]$ \cite[chapter 13]{cover2006}, even though the limits of the intervals in $Y$ are different from those considered in $X$, due to the shift caused by noise. Thus, assuming that there are $d=2^m$ quantization intervals in the unit interval, that the probability of occurrence is the same for all intervals, and that the bits in the binary expansion are equiprobable, references \cite{MicaelFMA:2024, ravilla2025arithmetic} show that that there is a symmetry between the quantization efficiency of direct and reverse reconciliation. Thus, their maximum values are given by:

\begin{equation}
	\label{eq:beta_max_direta}
	\beta_{q\ max }^{\rightarrow} = \frac{\sum_{i=1}^{m}C_i}{I(X;Y)} = \beta_{q\ max}^{\leftarrow}.
\end{equation}

Assuming that in real systems the channel capacity is not fully utilized and the individual error-correcting codes of each subchannel have rate $R_i \leq C_i$, and that the quantum channel can be well represented by a Gaussian channel \cite{jouguet2011long}, the efficiency over the entire reconciliation system, considering sub-optimal codes, is:

\begin{equation}
	\beta = \frac{\sum_{i=1}^{m}R_i}{I(X;Y)}.
\end{equation}

Since the code efficiency $\beta_{c} = \frac{R_i}{C_i}$ is seen as the ratio between the rate of real codes and ideal ones, $\beta$ can be rewritten as:

\begin{equation}
	\beta = \frac{\sum_{i=1}^m \beta_{c}C_i}{I(X;Y)} = \beta_{c} \beta_{q}.
\end{equation}
%

Thus, reconciliation efficiency depends directly on the rate of the available codes $R_i$ and how close they are to the capacity of each channel.

\section{Results}
\label{sec:results}

\begin{figure} 
	\centering
	\includegraphics[scale=0.7]{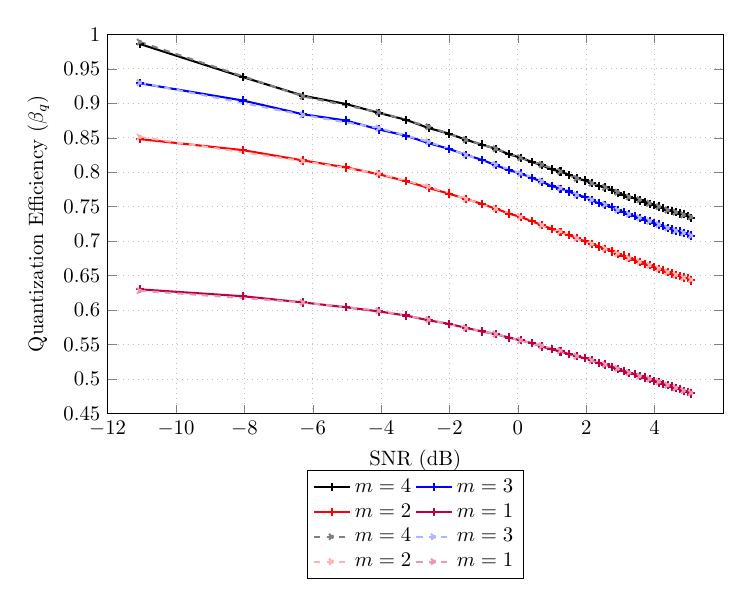}
	\caption{Reconciliation efficiency between $X$ and $V_i$ (solid lines) compared to reconciliation efficiency between $U$ and $V_i$ (dashed lines), considering $m$ subchannels.}
		\label{fig:Beta2}
\end{figure}

We performed simulations to verify the reconciliation efficiency in SNR regions between $-14$ and $2$ dB. Reference \cite{ravilla2025arithmetic} shows mutual information and efficiency results for $m \in \{1,...,7\}$, considering BSC channels, and, as in \cite{Laryssa:SbRT2018}, it shows that above 4 quantization bits, the improvement achieved in mutual information and quantization efficiency was almost negligible. Therefore, Fig. \ref{fig:Beta2} only shows reconciliation efficiency for $m \in \{1,...,4\}$.


It is worth noting that the calculation of mutual information \(I(X;V_i)\) comprises a continuous-discrete estimation, that is, for induced BIAWGN channels, whose results are generally obtained from estimators. In this work, we used the mutual information estimator based on reference \cite{kozachenko1987sample}, as used in reference \cite{MicaelFMA:2024}. The calculation of efficiency considering the mutual information between Alice and Bob's binary sequences (induced BSC channels) tends to present lower results, as shown in \cite{ravilla2025arithmetic}, since the quantization process by a low number of channels generates information losses.

Fig. \ref{fig:Beta2} shows that DTE performance improves with SNR decay, corroborating what was presented in \cite{MicaelFMA:2024}, making it a promising technique for applications in CVQKD reconciliation. It is also important to note that QKD protocols do not aim to extract information or generate keys identical to those transmitted at the beginning of the protocol, but rather to generate and distribute a secret key that must be unique (equal) between the two parties involved. The results obtained with quantization efficiency shows that the technique allows the distillation of secret keys, providing greater mutual information between Alice and Bob than the accessible information that Eva can obtain about the secret key.

Another interesting result was obtained by calculating the mutual information between the cumulative distribution function of \(X\), that is, \(U\), and the binary sequences of Bob's subchannels. From \(V_i\): \(I(U; V_i)\), the reconciliation efficiency with \(m \in \{1, \dots, 4\}\) channels was also calculated, and the values obtained were very close to those calculated previously, as can be seen in Fig. \ref{fig:Beta2}.

We also performed simulations in the error correction stage of AR, using a LDPC code with rate $R=1/4$, from the DVBS2 standard, used in satellite communications, with size $N=64800$, applied to the binary sequences obtained with the first quantization channel: $U_1$ and $V_1$. We run the reconciliation algorithm 250 times to obtain average values, using the modified sum-product algorithm for syndrome decoding, introduced by Liveris, Xiong, and Georghiades in \cite{van2004reconciliation}. We adopted the following stopping criteria: equality between the syndromes $S(U_1)$ and $S(V_1)$ or when the decoding algorithm reached 50 iterations.

\begin{figure}
	\centering
	\includegraphics[scale=0.7]{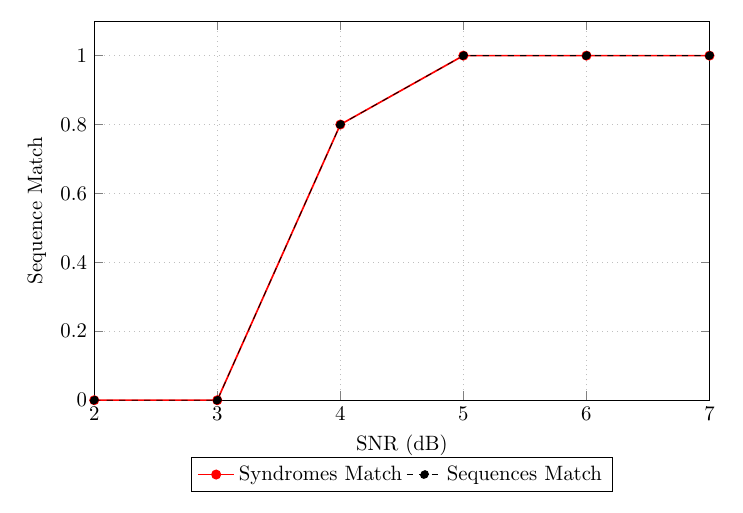}
	\caption{Syndromes match.}
		\label{fig:success}
\end{figure}

According to the calculated capacity for the quantization channels, the code rate only exceeds the capacity of channel 1 by about 2 dB ($C \approx 0.251$ bits/channel use). Fig. \ref{fig:success} shows that the syndromes begin to equalize at 3.8 dB, when the code rate already exceeds the channel capacity, showing the inefficiency of the code, and a distance of approximately 1.8 dB from the Shannon limit. This distance from the Shannon limit is due to two main factors: his theorem shows that error-free communication occurs when using asymptotically long codes, and the code used here is “short” in length compared to those recommended for this type of application (in the order of $N=10^5$); and the fact that the code used is not specifically designed for syndrome coding \cite{Plymouth_Syndrome}.

\begin{figure}
	\centering
	\includegraphics[scale=0.7]{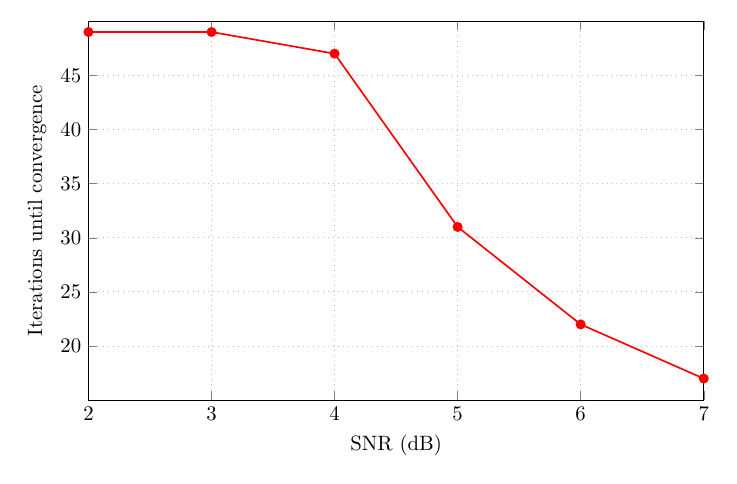}
	\caption{Iterations until convergence.}
		\label{fig:iterations}
\end{figure}

Nevertheless, whenever the syndromes were equal, the sequences were also equal, attesting to the effectiveness of applying LDPC codes for source compression (syndrome coding), based on the “bins” scheme presented by Slepian-Wolf \cite{Slepian_Wolf_1973, Liveris_2002}. Fig. \ref{fig:iterations} shows that even after 4 dB, the number of iterations required for convergence declines rapidly.

Simulations have shown that it is possible to apply AR in the reconciliation of CVQKD protocols, and encourage the application of more efficient and specific codes for syndrome coding — as these codes provide a higher error rate \cite{Plymouth_Syndrome} — in low SNR regions. The technique also provides greater versatility, as it allows multilevel coding and multi-stage decoding (MLC-MSD) for the transmission of other quantization channels, using adaptive rate codes or specific rates for each one.

\section{Conclusion}
\label{sec:conclusion}

In this paper we have presented some features of the AR protocol. Specifically, we have evaluated its efficiency in a range of -14 to 2 dB. Interestingly, simulations have shown that its efficiency increases as SNR diminishes. Moreover, we have also presented some of its advantages, such as low complexity. Additionally, we observed, by means of simulation, that it may be applied with syndrome coding, considering a context analogous to that of correlated sources, achieving perfect match in secret key acquisition above a SNR of 5 dB. Despite the fact that this is a high SNR in CVQKD scenario, we obtained this result using a code with a high rate and short length for this type of application. Thus, together, these results corroborates with the idea that AR scheme is a feasible and promising estrategy in CVQKD applications. As future work, further investigation may use longer length LDPC codes and lower SNR scenarios with AR applications. In addition, we will seek to obtain security proofs for the protocol against individual and collective attacks.

\appendices

\section{Evaluation of bit error probabilities}
\label{appendix:ber}

In this section, we evaluate the performance of the first and the second binary symmetric channels induced by our CVQKD procedure. 

\subsection{The first BSC performance}

Now we consider the case $j=1$. Firstly, notice that the bit error probability is given as
\begin{align}
\alpha_{1} & =  \Pr[V_{1} \neq U_{1}] \label{e:grund7.1} \\
& = 	\Pr[V_{1} = 1, U_{1} = 0] + \Pr[V_{1} = 0, U_{1} = 1] \label{e:grund7.2}  \\
& \overset{(a)}{=} 2 \Pr[V_{1} = 1, U_{1} = 0] \label{e:grund7.3}
\end{align}
where (a) happens due to symmetry.

We define the event 
\begin{equation}
D = \{U_{1} = 0, V_{1} = 1\} \nonumber 
\end{equation}
which is equivalent to
\begin{align}
D &\equiv  \{F_X(X)\leq 1/2, F_Y(Y)>1/2\} \nonumber \\
&= \{X<0, Y>0\} \nonumber \\
&= \{X<0, X+\sqrt{1/SNR}Z >0\} \nonumber \\
&= \{X<0, Z> -\sqrt{SNR}X\}. 
\end{align}

\begin{figure}
	\begin{center}
		\includegraphics[width=0.5\textwidth]{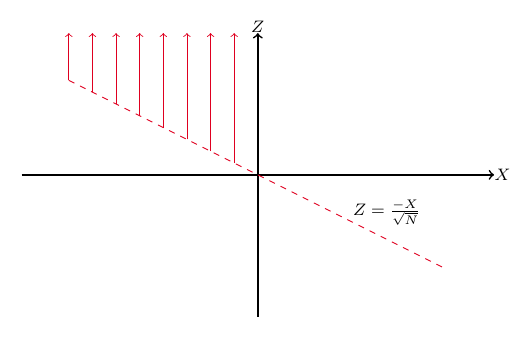}
		\caption{Region of integration for calculation of $\alpha_{1}$.}
		\label{fig:Regions_UXVY_2}
	\end{center}
\end{figure}

It is crucial at this point to note that both $F_X(X)$ as $F_Y(Y)$ must be interpreted as random variables themselves. We are now able to calculate $\alpha_{1}$ as follows:
\begin{align}
&\alpha_{1} = 2\mbox{Pr}[D] \nonumber \\
&= 2\int_{-\infty}^0\int _{-\sqrt{SNR}x}^{\infty} \dfrac{1}{2\pi} \exp\left\{-\dfrac{x^2+z^2}{2}\right\}dzdx \nonumber \\
&= 2\int_{-\infty}^0\int _{-\sqrt{SNR}x}^{\infty} \dfrac{1}{\sqrt{2\pi}} \exp\left\{-\dfrac{z^2}{2}\right\}dz\cdot
\dfrac{1}{\sqrt{2\pi}} \exp\left\{-\dfrac{x^2}{2}\right\}dx \nonumber \\
&= 2\int_{0}^{\infty} Q\Biggl(\sqrt{SNR}v\Biggr)\frac{1}{\sqrt{2\pi}}\exp\{-v^2/2\} dv, \label{e:grund9.3}
\end{align}
where in (\ref{e:grund9.3}) we have made the change of variable $v = -x$. The region of integration is illustrated in Fig. \ref{fig:Regions_UXVY_2}. This integral can be numerically calculated. 

\subsection{The second BSC performance}

Now we will specialize to the second BSC case and follow the reasoning of equations (\ref{e:grund9.3}), except by the limits of intervals. 

The following remark will be useful soon. Consider the numbers $2^{-i}, \ i = 1, 2, \ldots $, let us compute the reals
$
y_i = F_Y^{-1}(2^{-i}) 
$
in terms of the standard normal distribution.
\begin{align}
	2^{-i} & =  \int_{-\infty}^{y_i} \frac{1}{\sqrt{2\pi\left(\frac{1}{SNR}+1\right)}}  
	e^{-\frac{1}{2}\frac{u^2}{\frac{1}{SNR}+1}} du \nonumber \\
	& =  
	\int_{-\infty}^{y_i/\sqrt{\frac{1}{SNR}+1}} \frac{1}{\sqrt{2\pi}} e^{-\frac{v^2}{2}} dv \nonumber \\
	& =  \Phi\left(\frac{y_i}{\sqrt{\frac{1}{SNR}+1}}\right)  \nonumber \\
	& =  \Phi\Biggl(\frac{F_Y^{-1}(2^{-i})}{\sqrt{\frac{1}{SNR}+1}}\Biggr)
\end{align}

From the last equality, inverting the the distribution function $\Phi()$  we obtain
\begin{align} 
y_i &  =   F_Y^{-1}\left(2^{-i}\right) = \sqrt{1 + \frac{1}{SNR}}\Phi^{-1}\Bigl(2^{-i}\Bigr) \nonumber \\
& = \sqrt{1 + \frac{1}{\textnormal{SNR}}}\Phi^{-1}\Bigl(2^{-i}\Bigr) \nonumber \\
& =  \sqrt{1 + \frac{1}{\textnormal{SNR}}}x_i. \label{e:grundB.3Events.3}
\end{align}
where, for ease the notation we set $x_i = \Phi^{-1}\Bigl(2^{-i}\Bigr)$. 

\begin{figure}[htpb]
	\centering
	\includegraphics{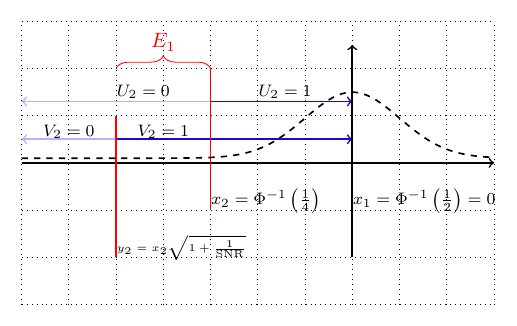}
	\caption{Illustration of the event of error $E_1$, according to the values for $U_2$ and $V_2$ in real axis.\label{fig:RegionsUXVY_1}}
\end{figure}

Now, let us define the event $E =\{U_2 = 0 , V_2 =1\}$. Fig. \ref{fig:RegionsUXVY_1} illustrates one manner by which event $E$ may happen.

\begin{figure}[htpb]
	\centering
	\includegraphics{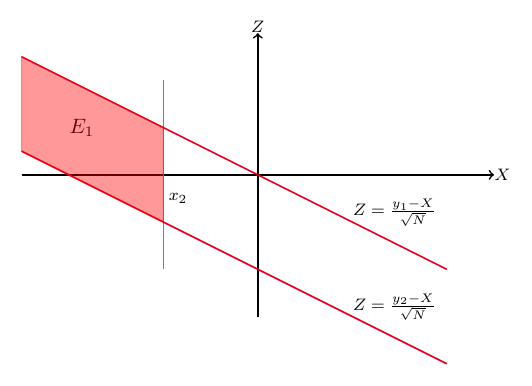}
	\caption{Integration region for event $E_1$.\label{fig:RegionsUXVY_3}}
\end{figure}

Event $E$ can happen in four mutually exclusive ways:
\begin{align}
E_1 &= \{F_X(X)<1/4 , 1/4<F_Y(Y)< 1/2\} \nonumber \\ 
E_2 &= \{F_X(X)<1/4 , F_Y(Y)> 3/4\} \nonumber \\
E_3 &= \{1/2<F_X(X)<3/4 , 1/4 < F_Y(Y) <1/2\} \nonumber \\
E_4 &= \{1/2<F_X(X)<3/4 , F_Y(Y) > 3/4\} \nonumber \\
E &= \cup_{i=1}^4 E_i, \nonumber 
\end{align}
and we will describe them in terms of the variables $X$ and $Z$, considering the correct limits of integration of the bivariate normal density $f_{XZ}$, as shown next (Fig. \ref{fig:RegionsUXVY_3} illustrates the region of event $E_1$ in the $XZ$ plane).
\begin{align}
    \Pr[E_1] &= \Pr [X<\Phi ^{-1}(1/4), F_Y^{-1}(1/4)<Y<F_Y^{-1}(1/2)] \nonumber \\
    &= \Pr \left[X<x_2, y_2<X+\frac{1}{\sqrt{SNR}}Z<0 \right] \nonumber \\
    &= \Pr \left[X<x_2, \dfrac{y_2-X}{\sqrt{1/SNR}}<Z<-\dfrac{X}{\sqrt{1/SNR}} \right] \nonumber \\    
	\Pr[E_1]   &  \overset{(a)}{=} 
	\int_{-\infty}^{x_2} 
	\int_{\frac{y_2 - x}{\sqrt{1/SNR}}}^{\frac{- x}{\sqrt{1/SNR}}} \frac{1}{2\pi}e^{-\frac{x^2 + z^2}{2}} dzdx
	\label{e:grundB.5} \\
	&= 
	\int_{-\infty}^{x_2} \frac{1}{\sqrt{2\pi}}e^{-\frac{x^2}{2}} \Biggl(
	\Phi\left(\frac{-x}{\sqrt{1/SNR}}\right) - \nonumber\\
	& \Phi\left(\frac{y_2 -x}{\sqrt{1/SNR}}\right)
	\Biggr)dx \nonumber \\
&=	\int_{-\infty}^{x_2} \frac{1}{\sqrt{2\pi}}e^{-\frac{x^2}{2}} (
	\Phi(-\sqrt{SNR}x) - \Phi(\sqrt{SNR+1}x_2\nonumber \\
	& -\sqrt{SNR}x))dx \nonumber \\
\end{align}

Similarly, we may evaluate the probabilities associated with $E_2$, $E_3$ and $E_4$.
\begin{align}
\Pr[E_2] &= \Pr[F_X(X)<1/4, F_Y(Y)>3/4] \nonumber \\
&= \Pr[X<x_2, X+\frac{1}{\sqrt{SNR}}Z> F_Y^{-1}(3/4)] \nonumber \\
&= \Pr[X<x_2,Z >\sqrt{SNR+1}\Phi^{-1}(3/4)-\sqrt{SNR}X] \nonumber \\
&= \int _{-\infty}^{x_2}\int_{\sqrt{SNR+1}\Phi^{-1}(3/4)-\sqrt{SNR}x}^{+\infty}\dfrac{1}{2\pi}e^{-\frac{(x^2+z^2)}{2}}dz dx \nonumber \\
&= \int _{-\infty}^{x_2} \dfrac{1}{\sqrt{2\pi}}e^{-\frac{x^2}{2}}[1-\Phi(\sqrt{SNR+1}\Phi^{-1}(3/4)-\nonumber \\
&\sqrt{SNR}x)]dx \nonumber \\
\end{align}

\begin{align}
\Pr[E_3] &= Pr[1/2<F_X(X)<3/4, 1/4<F_Y(Y)<1/2] \nonumber \\
&= \Pr[x_1<X<\Phi^{-1}(3/4), \nonumber\\
& F_Y^{-1}(1/4)<X+\dfrac{1}{\sqrt{SNR}}Z< 0] \nonumber \\
&= \Pr[x_1<X<\Phi^{-1}(3/4), \nonumber \\
& \sqrt{SNR+1}x_2-\sqrt{SNR}X<Z<-\sqrt{SNR}X] \nonumber \\
&= \int _{0}^{\Phi^{-1}(3/4)}\int_{\sqrt{SNR+1}x_2-\sqrt{SNR}x}^{-\sqrt{SNR}x}\dfrac{1}{2\pi}e^{-\frac{(x^2+z^2)}{2}}dz dx \nonumber \\
&= \int _{0}^{\Phi^{-1}(3/4)} \dfrac{1}{\sqrt{2\pi}}e^{-\frac{x^2}{2}} [\Phi(-\sqrt{SNR}x)-\Phi(\sqrt{SNR+1}x_2 \nonumber \\
&-\sqrt{SNR}x)]dx \nonumber \\
\end{align}
and,
\begin{align}
\Pr[E_4] &= \Pr[1/2<F_X(X)<3/4, F_Y(Y)>3/4] \nonumber \\
&= Pr[x_1<X<\Phi^{-1}(3/4), \nonumber \\ 
& Z>-\sqrt{SNR}X+\sqrt{SNR+1}\Phi^{-1}(3/4)] \nonumber \\
&= \int _{0}^{\Phi^{-1}(3/4)}\int_{-\sqrt{SNR}x+\sqrt{SNR+1}\Phi^{-1}(3/4)}^{\infty}\dfrac{1}{2\pi}\cdot \nonumber \\
&\exp\left\{-\frac{(x^2+z^2)}{2}\right\}dz dx \nonumber \\
&=\int _{0}^{\Phi^{-1}(3/4)}[1-\Phi (-\sqrt{SNR}x+\nonumber \\
&\sqrt{SNR+1}\Phi^{-1}(3/4))]\dfrac{1}{\sqrt{2\pi}}\exp\left\{-\frac{x^2}{2}\right\}dx \nonumber \\
\end{align}


\bibliographystyle{ieeetran}
\bibliography{bibliography}

\end{document}